# A novel technique based on a cylindrical microwave resonator for high pressure phase equilibrium determination.

Rodrigo Susial[a], Ángel Gómez-Hernández[a], Daniel Lozano-Martín[a], Dolores del Campo[b], M. Carmen Martín[a], José J. Segovia[a,*]

[a]TERMOCAL (Thermodynamics and Calibration) Research Group, Research Institute on Bioeconomy, Universidad de Valladolid, Paseo del Cauce 59, 47011, Valladolid, Spain.

[b]Centro Español de Metrología, Alfar 2, 28760 Tres Cantos, Madrid, Spain

* To whom correspondence should be addressed (e-mail): jose.segovia@eii.uva.es

**Abstract:**

The development of a novel technique based on a cylindrical microwave resonator for high pressure phase equilibrium determination is described. Electric permittivity or dielectric constant is a physical property that depends on temperature and pressure $\varepsilon(p,T)$. Based on this property, a measuring technique consisting of a cylindrical resonant cavity that works in the microwave spectrum has been developed. Equilibrium data of fluid mixtures are measured at high pressure using a synthetic method, where phase transition is determined under isothermal conditions due to the change of the dielectric constant. This technique may be a more accurate alternative to conventional visual synthetic methods. The technique was validated measuring pure $CO_2$, and phase behaviour was then determined for two binary mixtures $\{CO_2 (0.6) + CH_4 (0.4)\}$ and $\{CO_2 (0.4) + CH_4 (0.6)\}$, results for which are presented. These systems are interesting for the study of biogas–like mixtures. In addition, data were compared with the equation of state used for natural gas GERG-2008, and also, they were modelled using Peng-Robinson equation of state and Wong-Sandler mixing rules, which are widely employed in chemical industries and which give good results.

**Keywords:**

Phase equilibria; High pressure; $CO_2$; $CH_4$; Biogas, Cylindrical microwave resonator; Equation of State.

1. **Introduction**

The need for accurate and reliable high-pressure phase equilibria data in several applications [1,2], couple with the search of new renewable energy sources such as biogas, has led to the development of novel experimental techniques to accurately measure the phase equilibrium properties of these new blends. Furthermore, experimental data are essential to develop new equations of state and models or to improve existing ones such as the GERG-2008 [3], which is applied to natural gas and other similar mixtures. Measuring high pressure phase equilibria is often the only available way to obtain reliable data due to the complexity and difficulty involved in predicting the behavior under these conditions [4].

This novel technique is based on a cylindrical resonant cavity in the microwave spectrum. Three models of microwave resonators were used for phase equilibria determination. The first was described by Rogers *et al.* [5] in 1985, based on a cylindrical resonant cavity with an evanescent waveguide [6,7]. Numerous authors [8-17] used a re-entrant resonator to determine high pressure phase equilibria. Recently, Underwood *et al.* [18] measured dew point of water using a quasi-spherical microwave resonator. Various methods have been used to obtain experimental data of high pressure phase equilibrium. The most widely used, however, and which accounted for 36.4 % of all published articles on high pressure phase equilibria between 2005 and 2008 [4], is the visual synthetic method with phase transition. The technique presented in this work aims to be more accurate, to involve less uncertainty and to offer an automated alternative to visual techniques [19]. The technique could be classified as synthetic method with phase transition detection and isothermal operation mode [20]. First, the technique was set up and checked by measuring the liquid-vapour saturation line of $CO_2$, before then measuring the phase behavior of mixtures related to biogas processes.

Biogas is primarily a gas mixture composed of $CO_2$ and $CH_4$ in addition to small amounts of other gases, which is produced by anaerobic decomposition of organic materials. There is interest in upgrading biogas to obtain so-called biomethane, due to its potential applications for injection into natural gas grid or its use as vehicle fuel. The $CO_2$ separation process from biogas is an intrinsic operation in biomethane production and, combined with carbon dioxide capture and storage (CCS) technology [21], there is bio-energy production with CCS (BECCS), with negative $CO_2$ emissions, thus contributing to climate change mitigation.

The European Commission has issued mandate M/475 to European Committee for Standardization (CEN) [22] concerning the specifications of biogas and biomethane for their injection into natural gas grids and use as transport fuel. Our research group was a partner in the "Metrology for Biogas" project EMRP ENG54 together with twelve European National Metrology Institutes where thermophysical property measurements were performed to characterize the thermodynamic behavior of biogas and biomethane [23-25].

## 2. Experimental

2.1 Materials

The pure $CO_2$ used to check the technique was supplied by Air Liquide in bottles with 4.95 MPa pressure, with a ≥ 99.7% purity in absolute volume and containing water impurities ≤ 200×10$^{-6}$ in volume.

One the other hand, ($CO_2$ +$CH_4$) mixtures were prepared in the Spanish National Institute of Metrology (CEM) using a gravimetric method following the ISO 6142:2011 standard for preparing calibration gas mixtures and were supplied in 5 dm$^3$ aluminium bottles. For these mixtures, pure carbon dioxide was supplied by Carburos Metálicos (Air Products) with a certified mole fraction purity of 0.99995, and pure methane was supplied by Praxair with a certified mole fraction purity of 0.999995. The composition of these mixtures, in mole fraction of $CO_2$, and their corresponding expanded uncertainties ($k$=2) are (0.601623 ± 0.000040) and (0.400637 ± 0.000024). The densities of these mixtures were previously measured in our laboratory by Mondéjar *et al.* [26] using a single sinker densimeter. Table 1 contains data of all the pure compounds.

Table 1. Material description.

| Chemical name | Source | Mole fraction purity[a] | Purification method |
|---|---|---|---|
| [b]$CO_2$ | Air-Liquide | ≥0.997 | None |
| [c]$CO_2$ | Air Products | ≥0.99995 | None |
| $CH_4$ | Praxair | ≥0.999995 | None |

[a] Given by the supplier using gas chromatography
[b] Used for checking the technique
[c] Used for the mixtures

2.2 Experimental technique

The physical principle of detection is based on the change of trend of dielectric constant which occurs due to phase transition and also involves a discontinuity in the resonance frequency of the cavity. The experimental technique developed is classified as a synthetic method with non-visual phase transition [27] and isothermal operation mode. Synthetic methods involve preparing a mixture of accurately known composition. The experimental procedure involves programming pressure ramps, starting with the sample in homogeneous phase until the appearance of a new phase (bubble or dew points). The diagram of the technique is shown in Figure 1 1 and the different elements are described below.

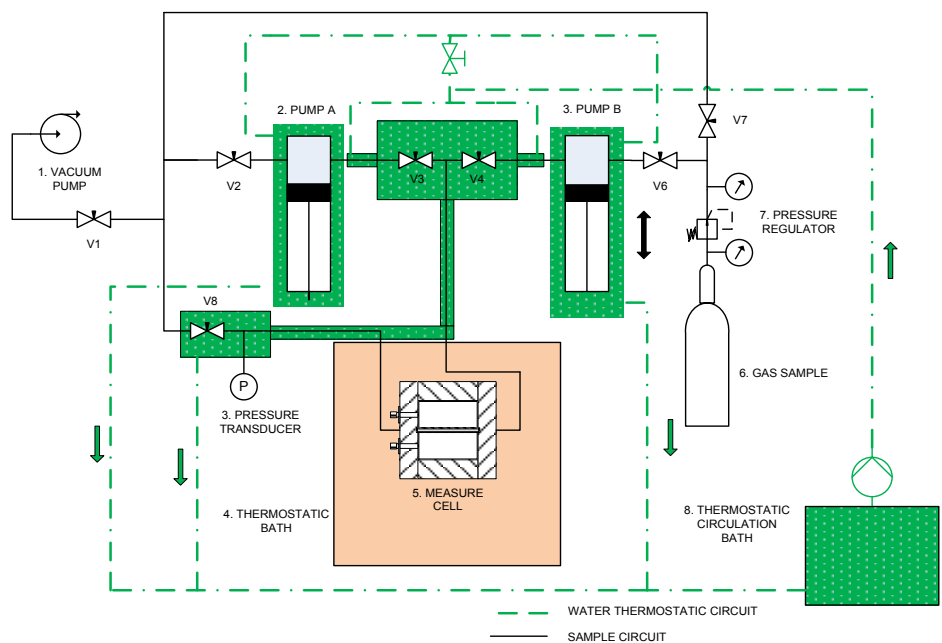

Fig 1. Diagram of the new experimental technique.

A precision thermostatic Hart Scientific bath (model 7340) is used to maintain the measuring cell at isothermal conditions. The bath can operate in a temperature range from -40°C to 150°C with a homogeneity of 6 mK and a stability of 5 mK. Two syringe pumps A and B Isco (model 260 D) work in a flow rate range from 1 µL/min to 90 mL/min at maximum pressure of 51.7 MPa. One of the pumps is programmed for pressure ramps. Additionally, an external circulation thermostatic bath (Julabo F-25) controls the temperature of the syringe pumps and sample circuit. When mixtures are measured, in order to homogenize them, one syringe pump works with positive displacement and another with negative displacement. An ASL F100 thermometer, with two Pt100 sensors, was used to measure temperature. These sensors are located into the bath at both sides of the measure cell. The thermometer was calibrated in the temperature range from 233.15 K to 338.15 K with an expanded uncertainty ($k = 2$) of 30 mK. For pressure measurements, a Druck DPI-145 calibrated pressure indicator with a PDCR 911-1756 external transducer was used. Its expanded uncertainty ($k = 2$) is 0.0073 MPa in a pressure range from 0 to 10 MPa. Both devices were calibrated in our laboratory using equipment traceable to national standards.

The measuring cell consists of a cylindrical resonant cavity that works in the microwave spectrum. The internal volume of the cavity is 98.17 cm$^3$, and a 15 cm long sapphire tube that contains the sample is located inside this cavity. This sapphire tube has an external diameter of 3.99 mm, an internal diameter of 2.01 mm and an internal volume of 0.52 cm$^3$. The cylindrical resonator was made of copper-zirconium (Luvata ZrK015) due to its low conductivity and mechanical simplicity. Figure 2 shows a picture of the resonant cylindrical cavity with the sapphire tube and a schematic view of the cell.

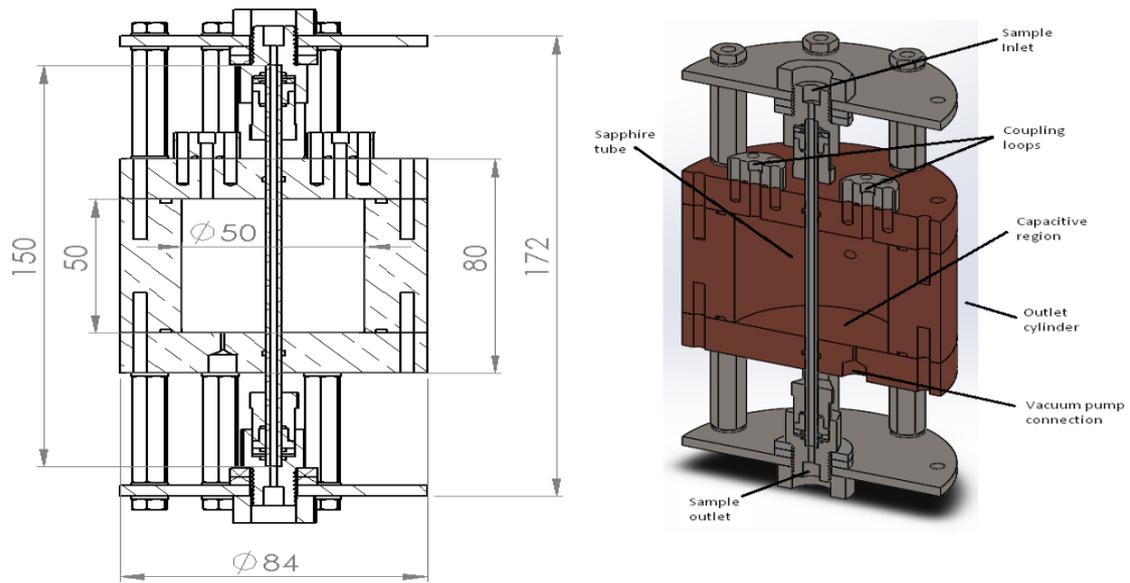

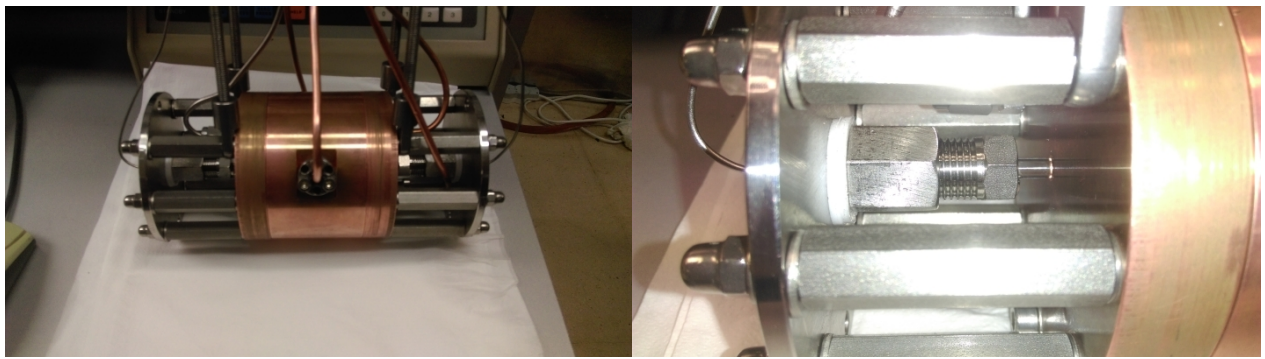

Fig 2. Schemes and pictures of the resonant cylindrical cavity with the sapphire tube of sample.

Two couplings made by electric probes are used to transfer microwave signal, using a vector network analyser (VNA model N5230C) connected to the cell using coaxial cables. The VNA was calibrated with an Agilent Calibration Kit 8510 and configured to measure the complex scattering coefficient $S_{21}$ through the cavity and connected to the antennas with GoldPt SMB r/a plg-plg RG316 waveguides. The time base of the VNA is linked to a rubidium standard frequency to improve the accuracy and the thermal stability of microwave frequency measurements, resulting in a standard uncertainty for frequency of 10 ppm which is negligible in comparison with the uncertainty of pressure or temperature measurements [28].

As regards the principle of measurement in microwave resonant cavities, the first transversal magnetic mode ($TM_{010}$) was chosen because of its high strength electric field in the axis of the cylindrical cavity. $TM_{010}$ mode frequency is about 4.23 GHz. This feature is necessary in order to improve sensitivity.

As noted above, the experimental procedure involves programming pressure ramps, starting with the sample in homogeneous phase until the appearance of a new phase (bubble and dew points). A phase transition means a discontinuous density change and also causes a dielectric constant discontinuity. The relationship between density (ρ) and static dielectric constant ($\varepsilon_r$) is determined by the Clausius–Mossotti relation for non-polar fluids (Eq. 1):

$$(\varepsilon_r - 1)/(\varepsilon_r + 2) = (\rho N_A \alpha)/(3 M \varepsilon_0) \qquad (1)$$

where $N_A$ is Avogadro's number, $\alpha$ is the molecular polarizability, $M$ is the molar mass of the substance and $\varepsilon_0$ is the permittivity of the vacuum.

The physical principle of operating the cylindrical microwave resonator for phase equilibrium data collection is qualitatively based on the perturbation method in microwave resonant cavities [29] used extensively in dielectric constant determination techniques [30,31]. The following equations are relative to the perturbation method applied to the cylindrical microwave resonator with the sample tube.

$$(f_{01} - f_{02})/f_{02} = A(V_s/V_c)(\varepsilon'_{r2} - \varepsilon'_{r1})/\varepsilon'_{r1} \qquad (2)$$

$$1/Q_2 - 1/Q_1 = B(V_s/V_c)\varepsilon''_{r2} \qquad (3)$$

The $f_{02}$ is the resonance frequency in homogenous phase, $f_{01}$ is the resonance frequency after phase transition, $\varepsilon'_{r2}$ and $\varepsilon'_{r1}$ are the real part dielectric constant in homogenous phase and after phase transition, and $V_s$ and $V_c$ are volumes of the sample and the cavity, respectively. A and B are parameters that characterise the resonant cavity and can be obtained by calibration using samples of known permitivity. The quality factors of the cavity ($Q_2$, $Q_1$) are not taken into account when non-polar substances are measured and dielectric losses ($\varepsilon''_{r2}$) are negligible.

Due to the appearance of a new phase in the dew or bubble point, dielectric constant discontinuity is detected by a discontinuity in the resonance frequency of the cavity (Eq. 2). An example of the behaviour of the resonance frequency will be shown in the next section where the validation of the technique is explained.

Accurate resonance frequency measurements are performed by means of a Vector Network Analyser (VNA) model PNL- N5230C from Agilent Technologies, with a frequency range from 300 kHz to 13.5 GHz and a maximum power from 3.98 mW to 12.59 mW. The VNA instrument measures complex transmission coefficient, $S_{21}$ (ratio of the voltage transmitted to the received voltage), which is vital to fitting resonance measurement data [32]. One model for an ideal resonator is the lumped-element series Resistor Inductor Capacitor (RLC) circuit (Figure 3), which was used for the resonant mode $TM_{010}$ of the resonant cavity [33].

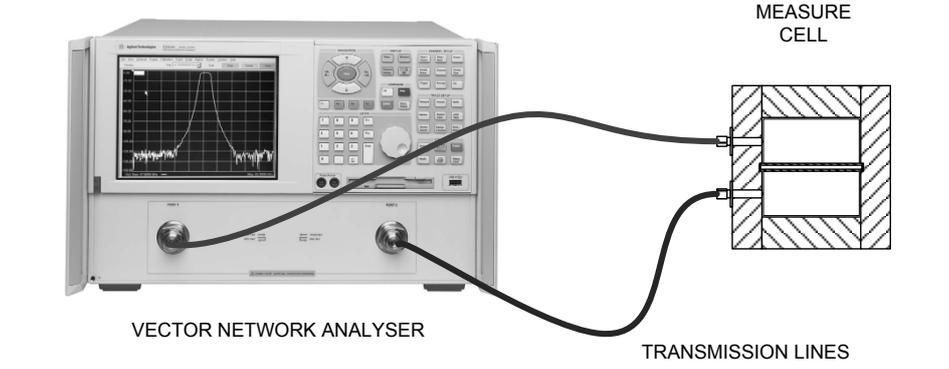

Figure 3. Equivalent circuit of microwave system.

There are different methods to obtain resonance frequency and quality factor in microwave resonators [32]. For this work, the complex transmission coefficient ($S_{21}$) as a function of frequency was fitted to a Lorentzian curve (Eq. 4).

$$|S_{21}(f)| = \frac{|\overline{S_{21}}|}{\sqrt{1 + 4Q^2\left(\frac{f}{f_0} - 1\right)^2}} \quad (4)$$

Here $|S_{21}(f)|$ is the module of complex transmission coefficient that depends on frequency, $|\overline{S_{21}}|$ is the maximum value, $Q$ is the quality factor, $f$ is frequency and $f_0$ is resonance frequency.

For better fitting results, a variable change was implemented and two sequential polynomic regressions are performed [34]. As an example, transmission coefficient as a function of pressure is shown in Figure 4, where experimental and fitted data are plotted.

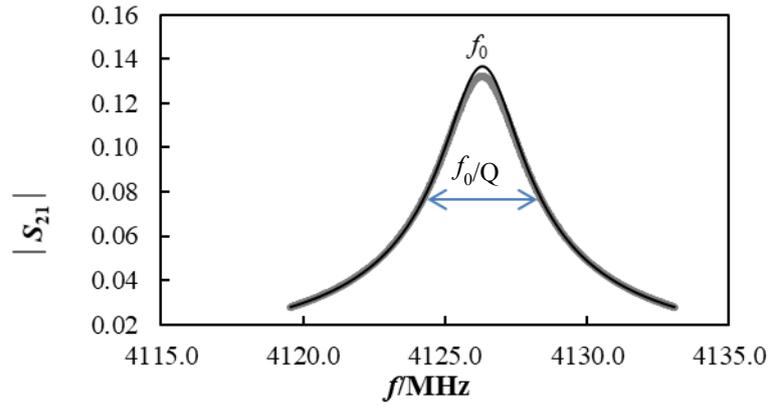

Figure 4. Example of the transmission coefficient as a function of frequency: (◆) Measured data (801 points) and Lorentzian fitted curve (─).

2.3 Experimental procedure

Before measurements are performed, the system is cleaned using a vacuum pump and then filled with the sample in homogenous phase conditions. During the isothermal experiment, the liquid or gas sample is maintained at constant temperature and pressure is modified until bubble or dew point occurs. The pressure ramp depends on sample volume, sample compressibility and the flow set point of the syringe pump, with its value varying in the range from $2\times10^{-5}$ MPa·s$^{-1}$ to $14\times10^{-5}$ MPa·s$^{-1}$. One of the syringe pumps is used for changing pressure. In addition, the pressure stability of the sample is also controlled by a thermostatic system in the surrounding area of the resonant cylindrical cavity near ambient temperature. Phase transition presents a change of trend in the dielectric constant and, therefore, in the resonant frequency of mode TM$_{010}$ of the cavity. When the mixtures are measured, pressure changes occur during phase transition. Homogenous phase was thus normalized by fitting the data to a second-degree polynomial in order to remove the pressure effect over dielectric constant in homogenous phase.

**3. Results**

3.1. $CO_2$ measurements

The experimental technique was validated by measuring $CO_2$ phase behaviour in the temperature range from 273.159 K to 304.044 K [21, 38] and comparing the results with those calculated using the Span and Wagner equation of state [39] through NIST REFPROP v 9.1 software [40]. An example of the experimental detection of $CO_2$ dew point is shown in Figure 5, where resonance frequency and pressure are plotted as a function of time for pure $CO_2$ at $T$ = 304.043 K. This isotherm is 86 mK below the critical point, where detecting changes proves challenging. However, as can be seen, the clear resonant frequency discontinuity means that phase transition occurs. It can be seen that between $t = 38$ to 41 min there is a noticeable change in the resonance frequency which changes from 4119.27 to 4120.54 MHz and the corresponding pressures are 7.359 MPa and 7.358 MPa, remaining constant within the uncertainty of the pressure measurement. In addition, certain non-linearity is observed prior to the jump of the frequency which can be attributed to the instability beginning of the condensation.

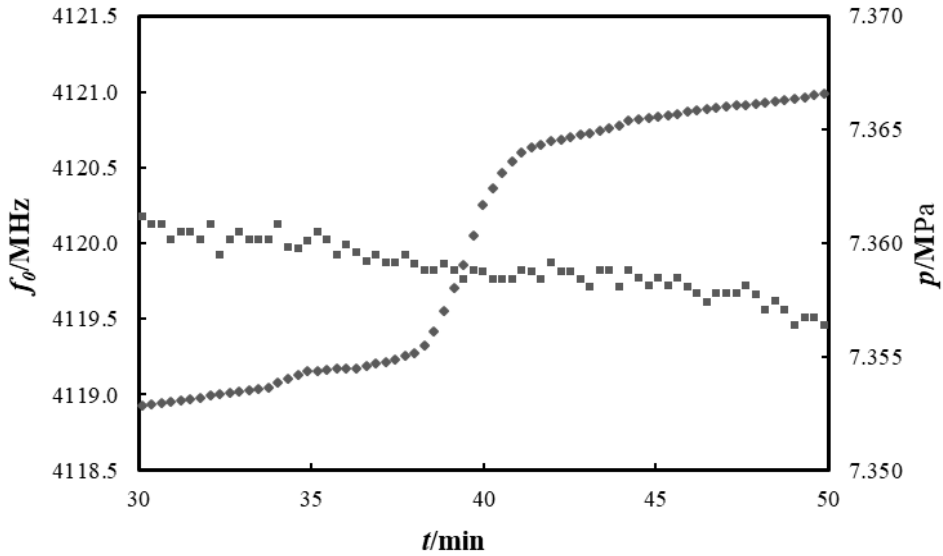

Figure 5. Experimental detection of $CO_2$ dew point at $T = 304.043$ K, (♦) resonance frequency and (■) pressure.

Table 2 contains phase equilibria data (pressure and temperature) measured with the new technique and the calculated pressures using Span and Wagner EoS [36] and relative deviations. The average absolute deviation is 0.002 MPa which is better than the standard uncertainty of the pressure measurement 0.004 MPa, with the average absolute relative deviation being 0.034%.

Table 2. Experimental $p$-$T$ data[a] of pure $CO_2$ and comparison with data calculated by Span and Wagner EoS [36]

| Set-up | $T_{exp}$/ K | $p_{exp}$/MPa | $p_{calc}$/MPa | $\delta p(\%)$[b] | Set-up | $T_{exp}$/ K | $p_{exp}$/MPa | $p_{calc}$/MPa | $\delta p(\%)$[b] |
|---|---|---|---|---|---|---|---|---|---|
| Dew | 273.165 | 3.4859 | 3.4865 | -0.017 | Bubble | 273.159 | 3.4857 | 3.4860 | -0.0086 |
| Dew | 278.198 | 3.9770 | 3.9743 | 0.068 | Bubble | 278.199 | 3.9750 | 3.9744 | 0.015 |
| Dew | 283.150 | 4.5005 | 4.5022 | -0.038 | Bubble | 283.150 | 4.5020 | 4.5022 | -0.0044 |
| Dew | 288.155 | 5.0905 | 5.0878 | 0.053 | Bubble | 288.155 | 5.0870 | 5.0878 | -0.016 |
| Dew | 293.140 | 5.7265 | 5.7277 | -0.021 | Bubble | 293.139 | 5.7264 | 5.7276 | -0.021 |

| | | | | | | | | |
|---|---|---|---|---|---|---|---|---|
| Dew | 298.151 | 6.4361 | 6.4344 | 0.026 | Bubble | 298.153 | 6.4301 | 6.4347 | -0.072 |
| Dew | 303.155 | 7.2144 | 7.2145 | -0.0014 | Bubble | 303.187 | 7.2128 | 7.2198 | -0.097 |
| Dew | 303.655 | 7.2974 | 7.2975 | -0.0014 | Bubble | 303.639 | 7.2909 | 7.2948 | -0.054 |
| Dew | 303.850 | 7.3290 | 7.3302 | -0.016 | Bubble | 303.850 | 7.3261 | 7.3302 | -0.056 |
| Dew | 304.043 | 7.3605 | 7.3628 | -0.031 | Bubble | 304.044 | 7.3588 | 7.3630 | -0.057 |

[a] Standard uncertainties: $u(T) = 0.015$ K; $u(p) = 0.004$ MPa

[b] $\delta p(\%) = 100 \cdot (p_{exp}/p_{calc} - 1)$

In addition, experimental data were compared with experimental data from literature: Kim and Kim [38] and Saleh and Wendland [39]. The relative deviations of experimental pressures in comparison with the values calculated using the reference equation of state [36] as a function of temperature are shown in Figure 6. It can be concluded that deviations in this work are similar to the data deviations in the literature. Near the critical point area, bubble point deviations are higher than dew point deviations in this work, although deviations are within expanded relative pressure uncertainty ($k=2$) of 0.1%.

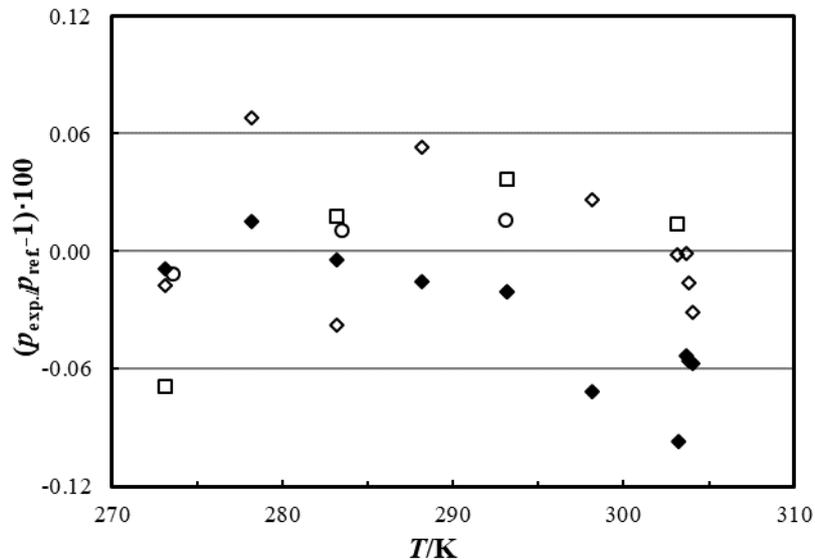

Fig 6. Relative deviations of experimental data to reference [36] of pure $CO_2$, (◆) bubble this work, (◇) dew this work, (□) experimental data from Kim and Kim [38], (○) experimental data from Saleh and Wendland [39].

3.2 $CO_2$ + $CH_4$ measurements

The {$CO_2$ (0.601623) + $CH_4$ (0.398377)} mixture was measured at the temperature range from 233.29 K to 270.21 K, near critical point and retrograde condensation area.

Experimental data were compared with calculated values using GERG-2008 [3] through NIST REFPROP v 9.1 software [37]. Results are summarized in Table 3. The recommended EA-4/02guide [40] was used for uncertainty calculations.

Table 3. Phase equilibria data[a] for the $CO_2$ (0.601623) + $CH_4$ (0.398377) mixture and comparison with data calculated using GERG-2008 EoS [3].

| Set-up | $T_{exp}$/ K | $p_{exp}$/MPa | $p_{EoS}$/MPa | $\delta p(\%)$[b] | Set-up | $T_{exp}$/ K | $p_{exp}$/MPa | $p_{EoS}$/MPa | $\delta p(\%)$[b] |
|---|---|---|---|---|---|---|---|---|---|
| Dew | 233.35 | 1.859 | 1.818 | 2.27 | Dew[c] | 271.12 | 8.314 | 8.350 | -0.43 |
| Dew | 243.35 | 2.687 | 2.656 | 1.17 | Dew[c] | 271.03 | 8.355 | 8.423 | -0.81 |
| Dew | 253.35 | 3.803 | 3.812 | -0.22 | Dew[c] | 270.93 | 8.463 | 8.481 | -0.22 |
| Dew | 258.34 | 4.486 | 4.555 | -1.52 | Dew[c] | 270.82 | 8.454 | 8.531 | -0.89 |
| Dew | 263.34 | 5.425 | 5.469 | -0.80 | Dew[c] | 270.33 | 8.687 | 8.673 | 0.17 |
| Dew | 266.32 | 6.143 | 6.140 | 0.05 | Dew[c] | 269.34 | 8.775 | 8.819 | -0.50 |
| Dew | 268.32 | 6.625 | 6.692 | -0.99 | Dew[c] | 268.32 | 8.907 | 8.898 | 0.11 |
| Dew | 269.34 | 7.011 | 7.023 | -0.17 | Dew[c] | 266.32 | 8.896 | 8.956 | -0.67 |
| Dew | 270.33 | 7.413 | 7.428 | -0.19 | Bubble | 263.34 | 8.872 | 8.909 | -0.41 |
| Dew | 270.82 | 7.712 | 7.961 | -3.12 | Bubble | 258.34 | 8.656 | 8.669 | -0.14 |
| Dew | 270.93 | 7.815 | 7.863 | -0.61 | Bubble | 253.35 | 8.268 | 8.327 | -0.71 |
| Dew | 271.03 | 7.857 | 7.779 | 1.01 | Bubble | 243.34 | 7.528 | 7.517 | 0.14 |
| Dew | 271.12 | 7.903 | 7.700 | 2.63 | Bubble | 233.35 | 6.702 | 6.667 | 0.54 |

[a] Standard uncertainties: $u(T) = 0.015$ K; $u(p) = 0.033$ MPa; $u(x_1) = 0.000020$

[b] $\delta p(\%) = 100\cdot(p_{exp}/p_{EoS}-1)$

[c] Retrograde dew point

A second {$CO_2$ (0.400637) + $CH_4$ (0.599363)} mixture was also studied in retrograde condensation area and near critical point, and experimental phase equilibria was obtained at $T$ = (233.223 to 248.19) K. Measured data were compared with the calculated values using GERG-2008 [3]. Both sets of data for this mixture are given in Table 4.

Table 4. Phase equilibria data[a] for the $CO_2$ (0.400637) + $CH_4$ (0.599363) mixture and comparison with data calculated using GERG-2008 [3]

| Set-up | $T_{exp}$/ K | $p_{exp}$/MPa | $p_{EoS}$/MPa | $\delta p(\%)$[b] | Set-up | $T_{exp}$/ K | $p_{exp}$/MPa | $p_{EoS}$/MPa | $\delta p(\%)$[b] |
|---|---|---|---|---|---|---|---|---|---|
| Dew | 233.35 | 3.051 | 3.036 | 0.49 | Dew[c] | 249.78 | 7.465 | 7.692 | -2.95 |
| Dew | 236.36 | 3.465 | 3.464 | 0.04 | Dew[c] | 249.74 | 7.555 | 7.529 | 0.35 |
| Dew | 238.36 | 3.818 | 3.784 | 0.88 | Dew[c] | 249.63 | 7.621 | 7.629 | -0.10 |
| Dew | 242.34 | 4.449 | 4.532 | -1.82 | Dew[c] | 249.53 | 7.682 | 7.692 | -0.13 |
| Dew | 245.35 | 5.228 | 5.244 | -0.30 | Dew[c] | 249.33 | 7.794 | 7.788 | 0.07 |
| Dew | 247.33 | 5.912 | 5.839 | 1.26 | Dew[c] | 248.33 | 8.065 | 8.030 | 0.44 |
| Dew | 248.34 | 6.296 | 6.216 | 1.30 | Dew[c] | 247.33 | 8.177 | 8.142 | 0.42 |
| Dew | 249.32 | 6.644 | 6.720 | -1.14 | Dew[c] | 242.34 | 8.180 | 8.185 | -0.07 |
| Dew | 249.53 | 6.906 | 6.871 | 0.52 | Bubble | 238.36 | 7.913 | 7.977 | -0.80 |
| Dew | 249.63 | 6.945 | 6.961 | -0.23 | Bubble | 236.36 | 7.839 | 7.831 | 0.11 |
| Dew | 249.74 | 7.068 | 7.090 | -0.32 | Bubble | 233.35 | 7.508 | 7.579 | -0.94 |
| Dew | 249.78 | 7.159 | 7.157 | 0.03 | | | | | |

[a] Standard uncertainties: $u(T) = 0.015$ K; $u(p) = 0.033$ MPa; $u(x_1) = 0.000012$

[b] $\delta p(\%) = 100 \cdot (p_{exp}/p_{calc} - 1)$

[c] Retrograde dew point

In addition, our experimental data were modelled using the phi–phi approach. For this purpose, the measured data were correlated with the Peng–Robinson equation of state (PR-EoS) [41] using the Wong-Sandler (WS) mixing rules [42]. Since an accurate description of VLE data is of great importance for the design and simulation of chemical processes, different equations of state have been used by chemical engineers for applications at moderate or high pressures [43], including the PR-EoS. Mixtures of substances, such as those used in this work, containing and/or associating polar components are of interest to chemical industries. Traditional thermodynamic models, such as cubic EoS, tend to perform well correlating VLE data. For this reason, a relative simple EoS model like PR [41] was used in this work (Eq. (5)).

$$p = \frac{RT}{v - b} - \frac{a(T)}{v(v + b) + b(v - b)} \qquad (5)$$

where $p$ is pressure, $T$ is temperature, $R$ is the universal gas constant and $v$ is the molar volume. Free volume effects and intermolecular attractive interactions are taken into account through parameters $b$ and $a$. For a pure component, the energy and size parameters are calculated, respectively, as follows

$$a(T) = 0.457235 \alpha(T) R^2 T_c^2 / p_c \qquad (6)$$

$$b = 0.077796 RT_c / p_c \qquad (7)$$

with $T_c$ and $p_c$ being the critical temperature and pressure of each pure compound, respectively. The correlation for the $\alpha(T)$ function is

$$\alpha(T) = \left[1 + \left(0.37464 + 1.5422\omega - 0.26992\omega^2\right)\left(1 - T_r^{0.5}\right)\right]^2 \qquad (8)$$

where $\omega$ and $T_r$ are, respectively, the acentric factor and the reduced temperature. The Wong–Sandler [42] (WS) mixing rules are given by

$$b_m = \frac{\sum_i^N \sum_j^N x_i x_j (b - a/(RT))_{ij}}{\left(1 - A_\infty^E/(CRT) - \sum_i^N x_i a_i/(RTb_i)\right)} \quad (9)$$

$$(b - a/(RT))_{ij} = \frac{1}{2}\left[(b - a/(RT))_i + (b - a/(RT))_j\right](1 - k_{ij}) \quad (10)$$

$$a_m = b_m \left(\sum_i^N x_i \frac{a_i}{b_i} + \frac{A_\infty^E}{C}\right) \quad (11)$$

where C is a numerical constant equal to $(\sqrt{2} - 1)/\sqrt{2} = -0.623$ and the $k_{ij}=k_{ji}$ is the binary interaction parameter for each system. In addition, $A_\infty^E$ is an excess Helmholtz free energy model at infinite pressure that can be equated to a low-pressure excess Gibbs energy model. In this study, we used the NRTL model [40] by,

$$\frac{A_\infty^E}{RT} = \sum_i^N x_i \frac{\sum_j^N x_j G_{ji} \tau_{ji}}{\sum_r^N x_r G_{ri}} \text{ with } \frac{G_{ij}}{RT} = exp(-\alpha_{ij}\tau_{ij}) \quad (12)$$

with $\alpha_{ij}=\alpha_{ji}$ as well as $\tau_{ij}$ and $\tau_{ji}$ being adjustable parameters, to obtain optimal agreement between theory and experiment for vapour–liquid equilibria. The Martín *et al.* [44] MATLAB program that uses the previous PR equations was applied to obtain the predictions of the systems of this work. The bubble and dew point routine of this program was used to obtain the optimal fitting parameters of this PR-EoS. The statistical parameter used was the objective function (OF), employing the liquid and vapour phase composition for minimization properties, as follows:

$$\text{Bubble point (OF)} = \sum_1^N (p^{exp} - p^{cal})_i^2 + \sum_1^N (y^{exp} - y^{cal})_i^2 \quad (13)$$

$$\text{Dew point (OF)} = \sum_1^N (p^{exp} - p^{cal})_i^2 + \sum_1^N (x^{exp} - x^{cal})_i^2 \quad (14)$$

The results obtained are reported in Table 5, and show that the binary interaction parameter provides a good representation of the mixture's actual behaviour. The mean absolute deviation of pressure (MAD) is also shown, and displays lower deviation compared with experimental data.

Table 5. Optimal parameters and mean absolute deviation (MAD)[a] of the experimental data using the Peng-Robinson [38] equation of state and Wong-Sandler [39] mixing rules (PR-WS EoS).

| $CO_2$ $(x)$+$CH_4$(1-$x$) | | $x = 0.601623$ | $x = 0.400637$ |
|---|---|---|---|
| PR-WS (EoS) | $k_{ij}$ | 0.2012 | 0.2211 |
| | $\alpha_{ij}$ | 0.3 | 0.3 |
| | $\tau_{ij}$ | 0.7473 | 1.5937 |
| | $\tau_{ji}$ | 0.3911 | -0.0490 |
| | MAD($p$)[a]/MPa | 0.066 | 0.080 |

[a] $\text{MAD}(p) = \sum_{i=1}^{n} |p_{\text{exp}} - p_{\text{calc}}|/(n-3)$

The adjustable parameters using the NRTL model provide good agreement between experimental and calculated data in both systems, obtaining average absolute deviations of 0.066 MPa and 0.080 MPa for the mixtures with a mole fraction of $CO_2$ of 0.6 and 0.4, respectively. When the data are modelled using Van der Waals mixing rules using only one parameter $k_{ij}$ for both mixtures, the MAD values increase slight to 0.085 MPa and 0.10 MPa, respectively. Both results are satisfactory taking into account that the expanded pressure uncertainty of ($k = 2$) is 0.066 MPa.

In order to compare our measurements with literature data, we only selected mixtures whose compositions are close to our measurements, with all the authors having measured vapour-liquid equilibria in isothermal conditions: Donnelly and Katz [45], Davalos *et al.* [46], Al-Sahhaf *et al.* [47], Wei *et al.* [48], and Webster and Kidnay [49]. The comparison was performed calculating the deviations of the experimental pressure from the calculated values using GERG-2008 [3].

For the {$CO_2$ (0.6)+$CH_4$ (0.4)} mixture, Figure 7 presents the pressure-temperature diagram showing the experimental data, and the calculated curves using GERG-2008 [3] and Peng-Robinson with Wong-Sandler mixing rules EoS [41,42]. In addition, deviations from GERG-2008 of our experimental data and those found in the literature [45-49] are shown in Figure 8.

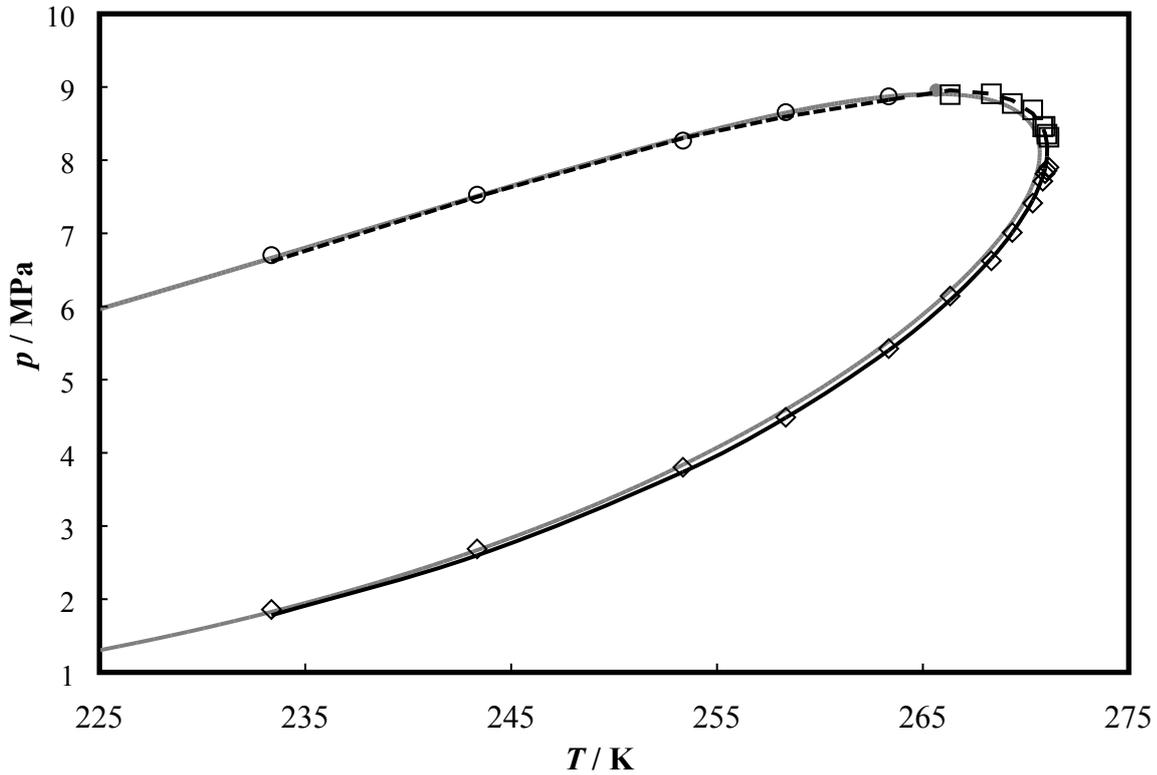

Figure 7. Phase behavior for the $CO_2$ (0.601623) + $CH_4$ (0.398377) mixture: (O) experimental bubble point, (□) experimental retrograde dew point, (◊) experimental dew point; (---) bubble point curve and (—) dew point curve calculated by GERG; (●) critical point GERG-2008 [3]; (---) bubble point curve and (—) dew point curve calculated by Peng-Robinson [41].

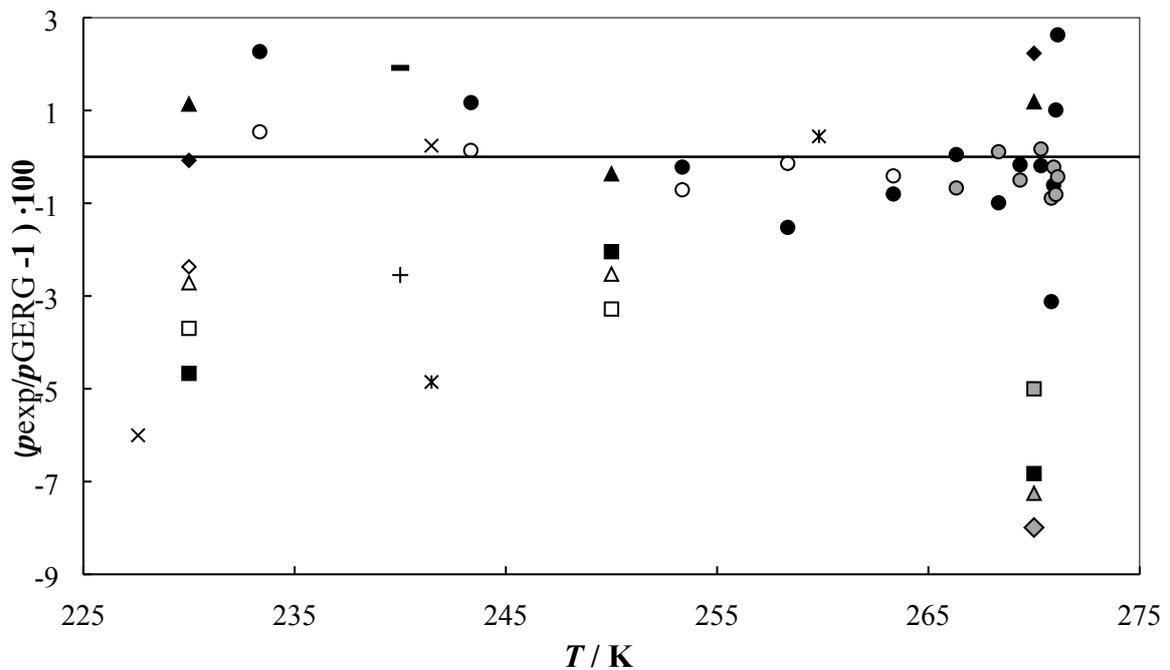

Figure 8. Relative deviations of experimental data from GERG calculations for the $CO_2$ (0.601623) + $CH_4$ (0.398377) mixture and comparison with literature. This work: ○ bubble point, ● dew point

(retrograde dew point in grey); Donnelly and Katz [45]: × bubble point, ∗ dew point; Davalos *et al.*[46]: □ bubble point, ■ dew point (retrograde dew point in grey); Al-Sahhaf *et al.* [47]: ‐ bubble point, ＋dew point; Wei *et al.* [48]: △ bubble point, ▲ dew point (retrograde dew point in grey); Webster and Kidnay [49]: ◇ bubble point, ◆ dew point (retrograde dew point in grey).

With regards to the {CO$_2$(0.4)+CH$_4$(0.6)} mixture, the phase envelope is plotted in Figure 9 where experimental data and the values calculated using GERG-2008 [3] and Peng-Robinson with Wong-Sandler mixing rules [41,42] are shown. Finally, deviations from GERG-2008 [3] of our experimental data and experimental data of literature [45,47-49] are depicted in Figure 10.

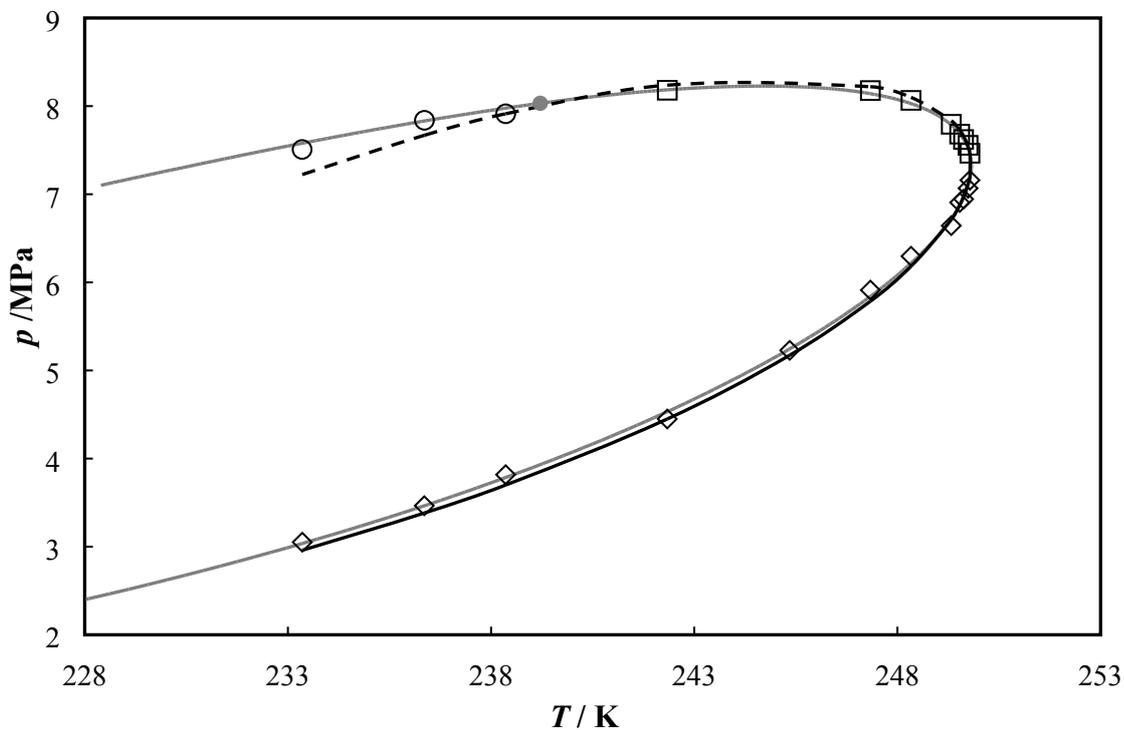

Figure 9. Phase behavior for the CO$_2$ (0.400637) + CH$_4$ (0.599363) mixture: (○) experimental bubble point, (□) experimental retrograde dew point, (◊) experimental dew point; (---) bubble point curve and (—) dew point curve calculated by GERG; (●) critical point GERG-2008 [3]; (---) bubble point curve and (—) dew point curve calculated by Peng-Robinson [41].

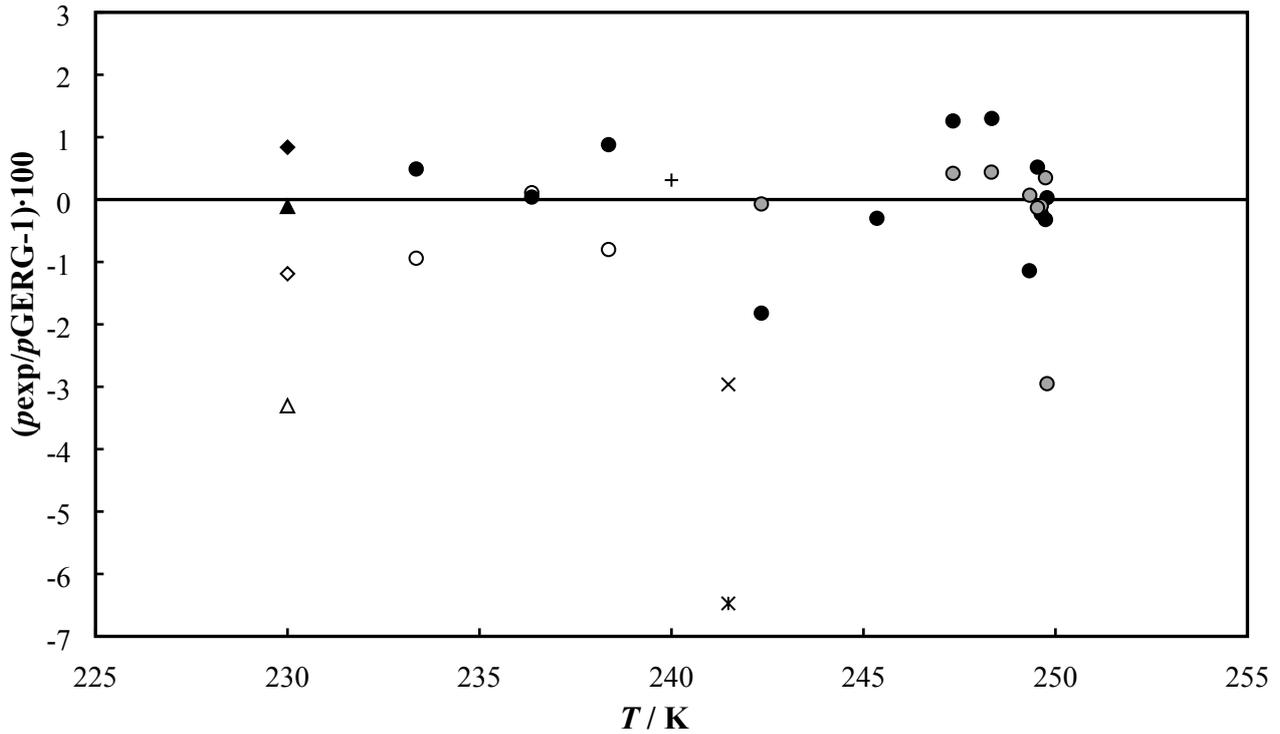

Fig 10. Relative deviations of experimental data from GERG calculations for the $CO_2$ (0.400637) + $CH_4$ (0.599363) mixture and comparison with literature. This work: ○ bubble point, ● dew point (retrograde dew point in grey); Donnelly and Katz [45]: × bubble point, ✲ dew point; Al-Sahhaf et al. [47]: + dew point; Wei et al. [48]: △ bubble point, ▲ dew point; Webster and Kidnay [49]: ◇ bubble point, ◆ dew point.

The comparison (Figure 8 and 10) shows the agreement between our experimental data and results using GERG-2008. Average absolute deviations are 0.034 MPa for the {$CO_2$ (0.6) + $CH_4$ (0.4)} mixture and 0.040 MPa for {$CO_2$ (0.4) + $CH_4$ (0.6)} mixture, with the average relative deviations being 0.60% and 0.67%, respectively, which are better than the literature values. Moreover, deviations in our measurements are below 3% even near the critical region, where measurements are quite difficult to perform as can be seen in Figure 8 where deviations range from 5% to 8% for literature data.

## 4. Conclusions

In this work, an accurate novel technique was developed to measure high pressure phase equilibrium by a synthetic method with phase transition based on a cylindrical microwave resonator with a sample tube. Pure $CO_2$ was used to validate the technique and results were good near critical point where more difficulties are encountered. Two binary ($CO_2$ + $CH_4$) mixtures were measured near critical point and retrograde condensation zone, and experimental data were compared with calculated data from the GERG-2008 reference equation and experimental data from the literature, obtaining good agreement. Finally, experimental data were also correlated with the PR-WS EoS, and good results were also obtained.

**Acknowledgments**

Financial support came from projects ENE2013-47812-R and ENE2017-88474-R from the Spanish Government and projects VA035U16 and VA280P18 from the Regional Government of Castilla y León. R. Susial is grateful for the support provided by the European Social Fund (ESF) and the Regional Government of Castilla y León.